\documentclass[apsrev,prx,twocolumn,amsmath,amssymb,longbibliography]{revtex4-1}% APS journal style
\usepackage{amsmath,amssymb,amsfonts,bm}
\usepackage{graphicx}
\usepackage{color}
\usepackage{bbold}
\usepackage{graphicx}% Include figure files
\usepackage{dcolumn}% Align table columns on decimal point
\usepackage{epstopdf}
\usepackage{color}
\usepackage{epsfig}
\usepackage{appendix}
\usepackage{mathdots}
\usepackage{titletoc}
\usepackage{url}
\usepackage{hyperref}
\usepackage{verbatim}
\newcommand{\lf}{\left(}
\newcommand{\ri}{\right)}

\newcommand{\im}{\operatorname{Im}}
\DeclareMathOperator{\Tr}{Tr}

\newcommand{\grad}{{\bf \bigtriangledown}}
\newcommand{\n}{\nonumber \\ }

\begin{document}
\title{Floquet Majorana zero and $\pi$ modes in planar Josephson junctions}
\author{Dillon T. Liu}
\email{dillon.liu@nyu.edu}
\author{Javad Shabani}
\author{Aditi Mitra}
\affiliation{Center for Quantum Phenomena, Department of Physics, New York University, New York, NY, 10003, USA}
\date{\today}

\begin{abstract}
We show how Floquet Majorana fermions may be experimentally realized by periodic driving of a solid-state platform. The system comprises a planar Josephson junction made of a proximitized heterostructure containing a 2D electron gas (2DEG) with Rashba spin-orbit coupling and Zeeman field. We map the sub-gap Andreev bound states of the junction to an effective one-dimensional Kitaev model with long range hopping and pairing terms. Using this effective model, we study the response of the system to periodic driving of the chemical potential, as applied via microwaves through a top-gate. We use the bulk Floquet topological invariants to characterize the system in terms of the number of zero and $\pi$ Majorana modes for experimentally realistic parameters. We present signatures of these modes on differential conductance and local density of states. A notable feature is subharmonic response to the drive when both Majorana
zero and $\pi$ modes are present, thus opening up the possibility of using this feature to clearly identify Majorana modes. We highlight the robustness of these features to disorder,
and show that despite hybridization due to long localization lengths, the main features of the subharmonic response are still identifiable in an experimental
system that is readily accessible.
\end{abstract}

\maketitle

\section{Introduction}\label{intro}
Non-equilibrium dynamical phenomena and topological phases of matter are two independent areas of study that have been a focus of intense
theoretical and experimental research in recent years. For the latter, a wide range of physical settings have been proposed, many focused on
realizing topological superconductivity \cite{AliceaRev, FlensbergMajorana, BeenakkerMajorana, StanescuMajorana, AguadoMajorana, LutchynMajorana}. This is primarily because
topological superconductors can host topologically protected qubits which are
tantalizing for quantum information purposes \cite{KitaevTQC, NayakRMP, Bravyi1, Bravyi2, Freedman}. Non-equilibrium phenomena on the other hand have
attained prominence recently because both theory and experiment can address fundamental questions about the nature of
thermalization in quantum systems, and the characterization of new dynamical phases that have no counterpart in thermal equilibrium \cite{AMreview, MoessnerFloquet}. Some examples of such states are time-crystals \cite{WilczekTC, OshikawaTC, KhemaniTC, YaoTC, LukinTC,MonroeTC} and non-thermal steady-states \cite{BlochMBL, MonroeMBL, FloquetMBL, KeyserlingkFloquet}.

We are concerned with a synthesis of two fields, topological superconductivity and Floquet periodic driving \cite{KitagawaFloquet,JiangFloquet,KitagawaWalk, RudnerFloquet, ReynosoFloquet, ThakurathiFloquet,KunduSeradjeh, NathanRudnerFloquet, FloquetLoss, RoyHarperFloquetHigherDim, RoyHarperTable}. Independently, these can give rise to distinct quantum states of matter, some summarized above, and together they may give rise to further exotic states of matter, such as Majorana fermions with ``time-crystal" subharmonic dynamics \cite{ElseFTC, ElsePrethermalTC}. The setting we choose for this synthesis is one that is well-characterized and has both practical and theoretical advantages \cite{Yacoby, Rokhinson, Marcus, ShabaniPRB2016, Henri, Marcus3, Fabrizio, Kaushini, Stanescu2018, Mayer}. In particular, we show how one may
realize Floquet Majorana fermions in a readily accessible and realistic experimental setting via planar Josephson junctions fabricated from semiconductor heterostructures proximity coupled to superconducting Al \cite{BergHalperin, Flensberg1, Flensberg2, YacobyPJJexpt, MarcusPJJexpt}.
\begin{figure}
	\includegraphics[width=0.47\textwidth]{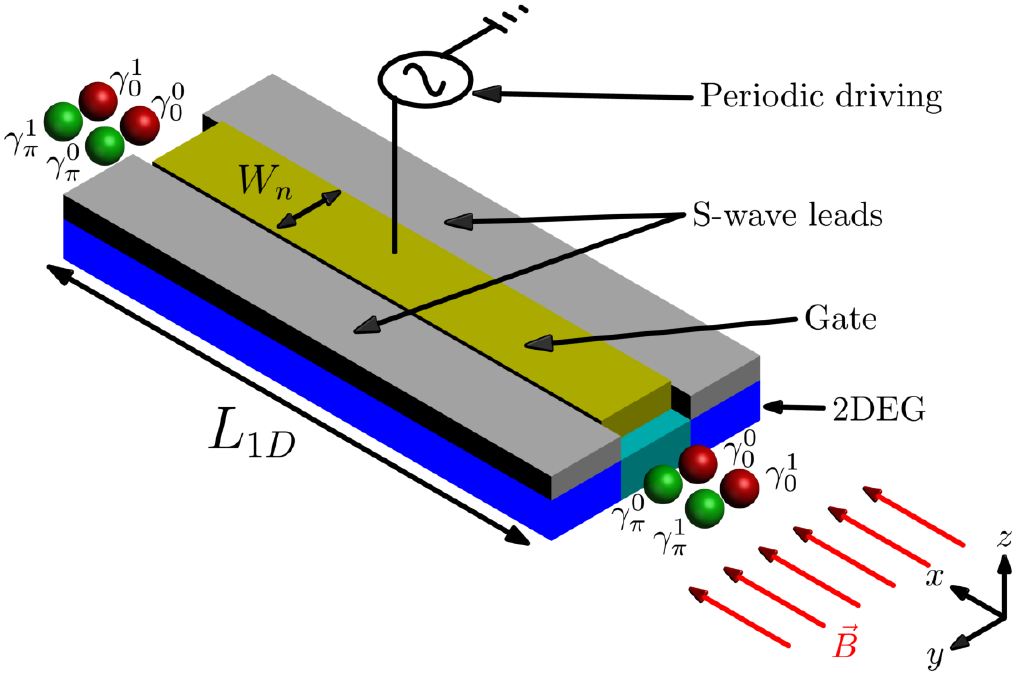}
	\caption{Proposed device to realize Floquet Majorana fermions based on Refs.~\onlinecite{BergHalperin, Flensberg1, Flensberg2, YacobyPJJexpt, MarcusPJJexpt}. A 2DEG with strong spin-orbit coupling is contacted to $s$-wave superconducting leads that have a phase difference $\phi$. An in-plane
longitudinal magnetic field, $\vec{B}$ is also applied. Floquet periodic driving is implemented by connecting the top gate to a microwave voltage source
to modulate the chemical potential. Floquet Majorana modes appear at the ends of the one dimensional (1D) normal channel and can be probed, e.g.,
with a quantum point contact or scanning tunneling probe. In this schematic, we have shown two Majorana zero modes and two Majorana $\pi$ modes at each end. As we discuss in the main text, generically the number of zero and $\pi$ modes is in $\left(\mathbb{Z},\mathbb{Z}\right)$. We will consider the case $(2,2)$ in this work.}
	\label{fig:illustration1}
\end{figure}

We note that these devices, in the absence of a periodic drive, are an active focus of current experimental efforts \cite{YacobyPJJexpt, MarcusPJJexpt}
and theoretical studies \cite{LiuShabaniMitra, HaimStern}.
We emphasize that using the ideas in Ref.~\onlinecite{LiuShabaniMitra}, the system here can be viewed as an effective one-dimensional model with generalized hopping and pairing
terms which go far beyond nearest neighbors, with the longer-ranged nature manifest in the resulting topological features of the system.
Long-range Kitaev chains have recently also been an
active area of theoretical interest \cite{NiuLRK, SenLRK,LeporiLRK, LeporiLRK2, MassiveEdgeModes, AlecceLRK, DuttaLRK, CaiLRK}. There are proposals for realizing these in physical systems, including schemes to induce long-range coupling via Floquet periodic driving of static nearest-neighbor Kitaev chains, ultracold atomic and trapped ion systems with photonic coupling, and unconventional solid-state systems \cite{FloquetLRK1, FloquetLRK2, Monroe, Zoller, Biercuk, PientkaLRK}. We show that one may realize these effective long range models in a straightforward solid-state setup as well.

By periodically driving such a semiconductor-superconductor (SNS) junction, as sketched in Fig.~\ref{fig:illustration1}, one may induce Floquet Majorana modes at the ends of a
quasi-one dimensional channel. As in the static situation, these modes may be probed by measuring the local density of states and differential conductance. Moreover, measurements of these quantities will show distinctive signatures indicating the presence of highly unusual dynamical modes and a subharmonic response to the drive, aka ``time crystal".

We anticipate that experiments of the type we presented here, involving Floquet driving, will be carried out in the near future. Further, we note that this platform has
tremendous tunability in the static case, so with the addition of Floquet driving, one has an extremely rich toolbox for studying diverse dynamical topological physics,
just one example of which is the effective generalized Floquet-Kitaev-type model that we have mapped the system to, and studied in detail here.

The paper is organized as follows. In Sec.~\ref{subgap}, we present the Hamiltonian used to model the experimental setup and explain related background,
such as the properties of the sub-gap Andreev bound states of the junction.
In Sec.~\ref{bulk_driving}, we study the bulk topological invariants for periodic driving using the Floquet formalism.
 This section also discusses the modification of the sub-gap Andreev bound states by periodic driving, and presents their quasienergy-phase relationship.
 In Sec.~\ref{wire_driving}, we map the complex periodically driven planar geometry to an effective model, namely a periodically driven finite wire
with long range pairing and hopping. This section includes
results for experimentally relevant quantities such as the conductance and the local density of states. Lastly, in Sec.~\ref{summary}, we provide a summary and outlook.
Some details have been relegated to the appendices.

\section{Planar Josephson junctions: Topological sub-gap states}\label{subgap}
We first briefly review the undriven system \cite{BergHalperin,Flensberg1, Flensberg2} on which our Floquet Majorana proposal builds on. Consider a 2DEG with strong spin-orbit coupling (e.g. InAs), an in-plane magnetic field, and $s$-wave superconducting leads that proximitize the 2DEG to leave a quasi one-dimensional channel. The Hamiltonian is given by ${\cal H} = \frac{1}{2}\Psi^\dag\cdot H_\text{BdG} \cdot \Psi$, where
\begin{align}
H_\text{BdG} &= \lf -\frac{\grad^2}{2m} - \mu\ri\tau_z - i\alpha a \lf\partial_x \sigma_y - \partial_y \sigma_x \ri \tau_z\n
&\quad\quad+ E_Z\lf y\ri \sigma_x + \Delta \lf y\ri\tau_+ + \Delta^\star\lf y\ri\tau_-,
\end{align}
and $\Psi = \lf c_{\uparrow}, c_{\downarrow},c^\dag_{\downarrow},-c^\dag_{\uparrow}\ri^{\intercal}$. $H_\text{BdG}$ includes Rashba spin-orbit coupling of strength $\alpha$, an in-plane Zeeman field $E_Z$ along the longitudinal direction, and two proximitized superconducting leads, characterized by an induced gap $\Delta$. $\sigma_i,\tau_i$ are the Pauli matrices acting in spin and particle-hole space respectively (with $\tau_\pm = \tau_x \pm i\tau_y$), and $a$ is a lattice spacing. In the transverse direction of the junction, we take
$\Delta \lf y\ri = \Delta e^{i\text{sgn}\lf y\ri\phi/2}\theta\lf \lvert y \rvert - W_n/2\ri$
and $E_Z\lf y\ri = E_{Z,s}\theta\lf\lvert y\rvert-W_n/2 \ri + E_{Z,n}\theta\lf W_n/2 - \lvert y\rvert\ri$.

We consider the bulk system by imposing periodic boundary conditions along the longitudinal $\hat{x}$ direction and open boundary conditions in the transverse $\hat{y}$ direction. Further, assuming translation invariance in the $\hat{x}$ direction, we can write the Hamiltonian in momentum space, $H_\text{BdG} \lf k_x, y \ri$ (we replace $k_x \to k$ below for notational efficiency). As has been previously discussed in Refs.~\onlinecite{BergHalperin, SatoEffectiveTSym}, $H_\text{BdG} \lf k \ri$ is in the Altland-Zinbauer \cite{AZ,KitaevClass, LudwigClass} symmetry class BDI (having particle-hole symmetry, an effective time-reversal, and chiral symmetry which
combines the two) and therefore has a topological invariant in $\mathbb{Z}$ \cite{KitaevBDI}.

\subsection{Sub-gap Andreev bound states}
The eigenstates of this bulk Hamiltonian include sub-gap Andreev bound states which have energy less than the induced gap in the superconducting
region. Thus the bound states are confined in the transverse direction, residing in the normal region. In the following, we consider a projected
Hamiltonian which includes only these bound states; in particular, we consider projecting to a single bound state band with Bogoliubov-de Gennes
dispersion $\pm E_0^{k}$,
\begin{align}
\tilde{H}_\text{BdG}^k &= E_0^{k} \lvert E_0^{k}\rangle \langle E_0^{k}\rvert - E_0^{k} \lvert-E_0^{k}\rangle \langle -E_0^{k}\rvert.
\end{align}

\begin{figure}
	\includegraphics[width=0.47\textwidth]{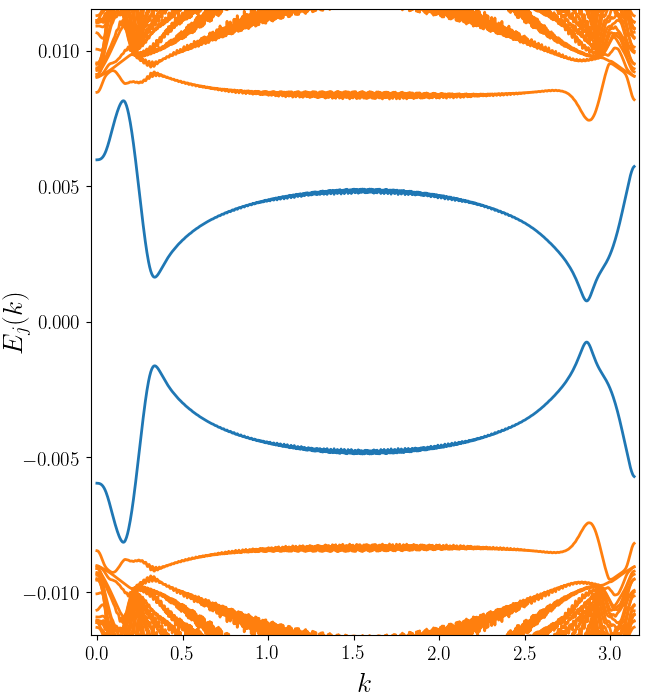}
	\caption{Bogoliubov-de Gennes spectrum. The lowest-lying state is highlighted in blue. They are energetically separated from modes in orange
which reside mostly outside the normal channel. We show only $k\in\left[0, \pi\right]$ because of the symmetry $k\to -k$.}
	\label{fig:static_spectrum}
\end{figure}

In Fig.~\ref{fig:static_spectrum}, the low-lying Bogoliubov-de Gennes spectrum of the full junction are shown and we highlight the lowest-lying sub-gap band. This is the band that we project to, ignoring the rest, all of which typically reside substantially outside the normal channel of the junction.

As the projected Hamiltonian retains all the symmetries of the full Hamiltonian, we can compute the topological invariant in a similar fashion. This invariant is computed by writing the Hamiltonian in a basis where the chiral symmetry operator is diagonal, and the Hamiltonian is block off-diagonal.
The determinant of the off-diagonal block may be used to define a complex phase, with the winding of the phase angle
being the topological invariant \cite{TewariSauInvariant}.

In the projected system, we similarly transform the Hamiltonian to the basis in which the chiral symmetry operator is diagonal, yielding the block off-diagonal form:
\begin{align}
\tilde{H}_\text{BdG}^k &=\begin{bmatrix}
    0 & \tilde{h}^k \\
    \tilde{h}^{k\dag} & 0
\end{bmatrix}.
\end{align}
Here, $\tilde{h}^k$ has only one non-zero eigenvalue, $\lambda^k_0$ because only one sub-gap band has been retained in the projection. Consequently, $\det\tilde{h}^k$ vanishes for all $k$. For this reason, we define a phase angle $\theta_k$ by considering the single non-zero eigenvalue. Defining,
$\alpha^k \equiv e^{i\theta^k} = \lambda^k_0/\lvert \lambda^k_0\rvert$, we extract the winding of this quantity. As in Ref.~\onlinecite{LiuShabaniMitra},
this winding angle can be used to define a 1D effective system for the projected sub-gap Hamiltonian. The topological invariant of this effective 1D system is,
\begin{align}
\nu &= \frac{1}{2\pi i} \int \frac{d\alpha^k}{\alpha^k}.
\end{align}
By the bulk-boundary correspondence, this invariant gives the number of zero-energy Majorana modes at the ends of a finite wire. For the numerical parameters that we pick below, the static system has winding number $\nu = 2$, indicating the presence of two topological edge modes.

\section{Floquet periodic driving} \label{bulk_driving}
\subsection{Floquet formalism and bulk Floquet topological invariants}
Periodically driven quantum systems can be fruitfully studied using the Floquet theorem formalism. According to Floquet theory, for systems with time-periodic Hamiltonians, $H\lf t + T\ri = H\lf t\ri$, the solutions to the Schr\"odinger equation can be written as $\lvert\psi\lf t\ri\rangle = e^{-i\epsilon_jt}\lvert\phi_j\lf t\ri\rangle$. In these solutions $\epsilon_j \in \left[-\Omega/2, \Omega/2 \right]$ are known as quasienergies and the Floquet modes are periodic in time $\lvert\phi_j\lf t\ri\rangle = \lvert\phi_j\lf t + T\ri\rangle$, where $\Omega = 2\pi/T$ and $\left[-\Omega/2, \Omega/2 \right]$ is the Floquet zone.

The Floquet modes are eigenstates for the time-evolution operator for a single period, $U\lf T+t, t\ri$, from which we can define $U\lf t^\star+T, t^\star\ri \equiv e^{-iH_F\lf t^\star\ri T}$, where $H_F\lf t^\star\ri$ is the Floquet Hamiltonian for a fixed starting point of the cycle, $t^\star$. Explicitly computing
$H_F\lf t^\star\ri$ is a significant challenge when working with periodically driven systems, however there are established methods such as the Sambe
approach \cite{Shirley, Sambe, Eckhardt} which we present below.

When written in terms of the Floquet solutions, the Schr\"odinger equation becomes
\begin{align}\label{FloquetSchrodingerEqn}
    \lf H\lf t\ri - i \partial_t\ri \lvert \phi_j\rangle &= \epsilon_j\lvert \phi_j\rangle.
\end{align}

The Floquet modes can be decomposed as $\lvert\phi_j\lf t\ri\rangle = \sum_m e^{-im\Omega t} \lvert \phi^m_j\rangle$ and the left-hand side of Eqn.~\ref{FloquetSchrodingerEqn} can be written as a block matrix in an extended Hilbert space known as Sambe space,
\begin{align}\label{sambespaceH}
H - i \partial_t &\equiv \begin{bmatrix}
    \ddots& \vdots & \vdots & \vdots & \iddots \\
    \cdots & H_{0}+\Omega & H_{1} & H_{2}  & \cdots \\
    \cdots & H_{-1} & H_{0} & H_{1} & \cdots \\
    \cdots & H_{-2} & H_{-1} & H_{0}-\Omega  & \cdots\\
    \iddots & \vdots & \vdots & \vdots  & \ddots
\end{bmatrix},\\
\left[H - i\partial_t \right]_{nm} &= \underbrace{\int_{t^\star}^{t^\star+T}\frac{dt'}{T} e^{-i\lf n-m\ri\Omega t'}H\lf t'\ri}_{\equiv H_{n-m}}  + \delta_{nm}m\Omega.
\end{align}

Terms $H_{\pm m}$ correspond to contributions from the $m$th harmonic of the time-dependent Hamiltonian. In the numerical simulations,
this matrix is truncated and diagonalized to obtain the Floquet quasienergy spectrum and the Floquet modes.

In order to compute $H_F\lf t^\star \ri$, we diagonalize the extended space operator in Eqn.~\eqref{sambespaceH} \cite{Eckhardt}.
The eigenvectors are used to construct the unitary
\begin{align}\label{sambespaceU}
\overline{U} &\equiv \begin{bmatrix}
    \ddots& \vdots & \vdots & \vdots & \iddots \\
    \cdots & U_{1,1} & U_{1,0} & U_{1,-1}  & \cdots \\
    \cdots & U_{0,1} & U_{0,0} & U_{0,-1} & \cdots \\
    \cdots & U_{-1,1} & U_{-1,0} & U_{-1,-1}  & \cdots\\
    \iddots & \vdots & \vdots & \vdots  & \ddots
\end{bmatrix},\\
%U_{n} &\equiv U_{m,m-n}.
\end{align}
%Consider $\overline{U}$ which diagonalizes the extended space representation of $H-i\partial_t$,
Thus,
\begin{align}
\overline{U}^\dagger \lf H-i\partial_t\ri \overline{U} &= \delta_{nm}\delta_{jj'} \lf\epsilon_j + m\Omega\ri.
\end{align}
Using the fact that $\overline{U}$ is translationally invariant in the extended space, so that
$U_{n} \equiv U_{m,m-n}$,
we can write $H_F\lf t^\star \ri = \hat{U}\lf t^\star \ri H_D \hat{U}^\dagger\lf t^\star \ri$, such that
\begin{align}
H_D &= \delta_{jj'} \epsilon_j,\\
\hat{U}\lf t\ri &=\sum_n e^{in\Omega t}U_n,\label{foldingeqn}
\end{align}
where $U_{n}$ are the matrices that comprise a column of the block matrix, $\overline{U}$. Eqn.~\ref{foldingeqn} indicates how the Floquet Sambe space is folded back into the physical space by summing over the extended space blocks.

In order to extract the topological aspects of Floquet systems, one has to revisit the notion of time-reversal invariance \cite{Delplace}.
Here, we consider periodic driving which has two points in the drive cycle denoted by $t^*$, which are time-reversal invariant in that $H(t^*-t)=H(t^*+t)$.
Combined with the usual particle-hole symmetry, we have a Floquet chiral symmetric system. The key to understanding the topological properties of
such a system is studying the Floquet Hamiltonian, $H_F\lf t^\star\ri$ with $t^\star$ one of the time-reversal invariant points in the cycle.

In these systems, there are two topological invariants, $\lf \nu_0, \nu_\pi\ri$, which are both in $\mathbb{Z}$.
These invariants can be computed using a procedure similar to the static case and involve evaluating a winding number from the off-diagonal
blocks of $H_F\lf t^\star\ri$. In the projected system, we again modify this procedure to extract the winding angle and compute the
Floquet invariants for the sub-gap system.

The corresponding topological Floquet modes that arise in finite systems differ from their static counterparts.
Instead of solely being pinned at zero energy, the topological modes may be pinned at zero or $\pm \Omega/2$ quasienergy.
We refer to these as Majorana zero modes (MZM) and Majorana $\pi$ modes (MPM). These two different modes can lead to qualitatively
new physics such as a modification of the
scaling of the entanglement entropy with a new definition for the central charge that now depends both on $\nu_0,\nu_{\pi}$ \cite{DanielPRL}.
Another effect is subharmonic behavior, a feature we discuss in detail below.

\subsection{Floquet topological Josephson junction}
The time-periodic driving we consider is a harmonic modulation of the chemical potential in the normal region of the junction.
Explicitly, this is given by
\begin{align}\label{drivingterm}
H_d^k \lf y, t\ri &= \sum_{\sigma,y\in \text{N}} \delta\mu \cos\lf \Omega t\ri c^\dag_{\sigma,k,y} c_{\sigma,k,y},
\end{align}
where $\delta\mu, \Omega$ are the amplitude and frequency of the drive, respectively, and $\text{N}$ is the set of sites
in the transverse direction, and restricted to the normal region of the junction.

We project the periodic driving operator to the eigenstate basis and truncate similarly to only the sub-gap bands. Explicitly we write
$\tilde{H}_d^k\lf t\ri$ in terms of $\lvert\pm E_0^{k}\rangle \langle\pm E_0^{k}\rvert$, i.e, the diagonal elements in
the projected basis, and in terms of $\lvert\mp E_0^{k}\rangle \langle\pm E_0^{k}\rvert$, the off-diagonal elements of
the projected basis. This projection neglects the coupling induced by the drive between the sub-gap states and the states above the gap.
It is a reasonable approximation to neglect this coupling for several reasons. First, the drive is spatially restricted to the normal channel
and the sub-gap states are the predominant eigenstates in that region. Second, the drive frequency is large compared to the static topological gap, but still smaller
than twice the induced gap from the superconducting proximity effect, $2\Delta$,
the energy required to break a Cooper pair. Third, the drive amplitude is taken to be small compared to all the other scales.

\begin{figure}
	\includegraphics[width=0.47\textwidth]{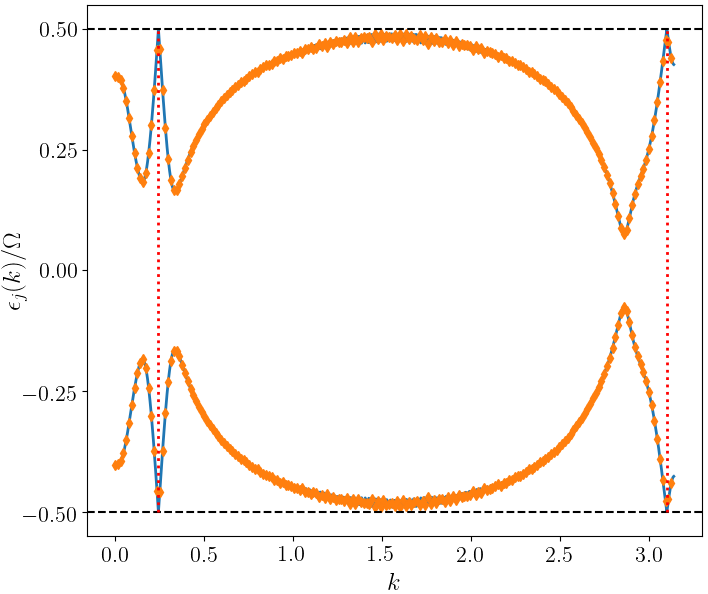}
	\caption{Quasienergy spectrum for periodically driven sub-gap states from the bulk model (blue lines). Dotted red lines correspond to $k$ points for which the drive frequency is resonant. Left resonant $k$ point centered on $0.244$. Right resonant $k$ point centered on $3.101$. The orange diamonds are the quasienergy spectrum from the periodically driven effective 1D problem described in Sec.~\ref{wire_driving}. We show only $k\in\left[0, \pi\right]$ because of the symmetry $k\to -k$. Horizontal dashed black lines correspond to $\epsilon_j = \pm \Omega/2$ and mark the edge of the Floquet zone.}
	\label{fig:quasienergy_spectrum}
\end{figure}

Applying the Floquet formalism described above, we now construct the Floquet Hamiltonian in Sambe space. Since the time-dependence of the driving in this case is given by a single harmonic, the only non-vanishing blocks are $H_0, H_{\pm1}$.  Then we diagonalize to find the quasienergy spectrum and the Floquet modes, e.g. in Fig.~\ref{fig:quasienergy_spectrum} the quasienergy spectrum is shown for a representative set of parameters. In units of the 2D hopping $t = 1/(2m^\star a^2)$,
the parameters chosen are $\mu = 0.003, \Delta = 0.01, E_{Z,n} = 0.02, E_{Z,s} = 0.001, \alpha = 0.1$, with $\phi = \pi/2, a \approx 20$nm, $m^\star = 0.03m_e$. We take the normal channel to have width $W_n = 10$ sites and the width of the leads is $150$ sites on each side in the transverse direction.
We consider a drive amplitude of $\delta\mu = 0.001$ and frequency $\Omega = 0.01$. In physical units, these values are $t\approx 16\text{meV}$, $\Delta \approx 160 \mu\text{eV}$,
$\Omega = \Delta \approx 38\text{GHz}$. Finally  the static topological gap, which
is the energy gap for the Andreev bound states, is $\Delta E_{g,0} \approx 13\mu\text{eV}$.
The Floquet topological invariants for this choice of parameters are $\lf \nu_0, \nu_\pi\ri = \lf 2, 2\ri$. Unless otherwise mentioned, we fix this choice of parameters in the discussions below to showcase the physics. We note that for our parameter choices $\Delta=\Omega$ which may lead to additional effects from higher harmonic coupling to above-gap bands which we neglect here. In an idealized case, one would have a separation of scales in which the topological gap is much smaller than the frequency, which is in turn much smaller than the induced superconducting gap: $\Delta E_{g,0} \ll \Omega \ll \Delta$.

\begin{figure}
	\includegraphics[width=0.47\textwidth]{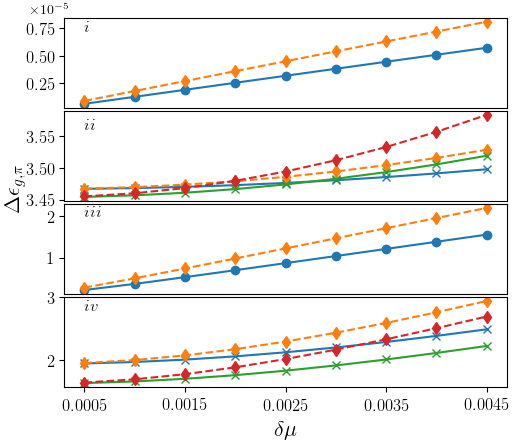}
	\caption{Dependence of the quasienergy $\pi$-gap, $\Delta \epsilon_{g,\pi}$, on the drive amplitude. We show the gap for several $k$
points near the resonances denoted in Fig.~\ref{fig:quasienergy_spectrum}. Solid lines are from the full tight-binding simulations, and dashed lines
correspond to estimates based on Appendix~\ref{RWA}. Each curve corresponds to the gap as a function of $\delta\mu$ for a fixed $k$ point.
(i) $k = 0.244$ is a resonant point and shows the linear dependence of the gap on the drive amplitude.
(ii) Blue and green ($\times$) curves correspond to $k$ points on either side of the resonance ($k=0.243,0.245$) and show a non-linear/quadratic dependence.
(iii) Again we see linear dependence, this time at the second resonant point, $k=3.101$. (iv) Lastly, we pick two $k$ points on either side of the
second resonance, $k=3.1,3.102$, and recover the deviations from linear dependence.}
	\label{fig:pi_gap_scaling}
\end{figure}

An important experimental consideration is the size of the quasienergy gaps around $0$ and $\pm\Omega/2$ in the quasienergy spectrum. The more familiar gap, $\Delta\epsilon_{g,0}$, is the analog of the usual static topological gap and is defined by the difference between $0$ and the closest $\epsilon_j$. We also define $\Delta \epsilon_{g,\pi}$, the so-called $\pi$-gap, by the difference between $\Omega/2$ and
the closest $\epsilon_j$. In Fig.~\ref{fig:quasienergy_spectrum}, one can see that while the gap around $\epsilon_j = 0$ is visible, $\approx 0.08$ (or $13\mu\text{eV}$ in physical units), the gap around $\pm \Omega/2$ is barely visible, $\approx 0.0015$, (or $0.3\mu\text{eV}$). This gap, however, is a function of the drive parameters. In particular, at the resonant point, it should depend on the drive amplitude in a linear fashion.

In Fig.~\ref{fig:pi_gap_scaling}, the scaling of the gap around $\pm \Omega/2$ is shown as a function of $\delta\mu$ and the linear dependence at the resonant $k$ point is clear.
This is useful for extrapolating the necessary amplitude in experiments to achieve sufficiently large gaps to resolve the MPMs.
We note that the size of these gaps can be estimated quite well by replacing the full Floquet formalism by a rotating wave approximation applied to a two-level system, with a general periodic driving. Further details of the estimation are given in Appendix~\ref{RWA}. Results here are presented for a zero gap and $\pi$ gap corresponding to $\Delta\epsilon_{g,0} \approx 0.08$ (or $13\mu\text{eV}$ in physical units) and $\Delta\epsilon_{g,0} \approx 0.0015$ ($0.3\mu\text{eV}$), respectively.

\subsection{Floquet Andreev bound state phase relation}
The Andreev bound states of the Josephson junction are expected to display a characteristic relationship between the energy and phase difference across the
junction. This dependence can give rise to distinctive experimental signatures and provides an additional perspective on the physics of the system.
Here we contrast the static and Floquet versions of the phase dependence in the bound state energy and quasienergy spectra.
\begin{figure}
	\includegraphics[width=0.5\textwidth]{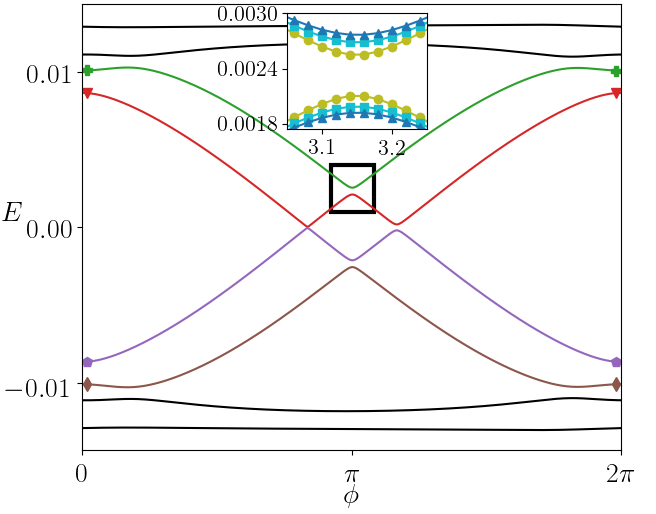}
	\caption{Andreev bound state spectrum for static system as a function of $\phi$. Characteristic behavior of a SNS junction is
seen here, with bound states showing $2\pi$ periodicity. Gap openings are generically present due to finite normal reflection processes at the
S-N boundaries. The four states with energy closest to zero are shown in color and with distinguishing markers at the edges. Inset: gap at $\phi=\pi$ for $\mu = 0.003, 0.005, 0.007$ shows that the gap increases with increasing chemical potential in this regime.}
	\label{fig:static_abs_ephi}
\end{figure}

In Fig.~\ref{fig:static_abs_ephi}, we present the BdG spectrum for a static junction as a function of $\phi$. The eight states shown are those which are closest to zero energy; of those eight states, we focus on the four with smallest energy magnitude (shown in color and with markers at the edges of the figure). The relationship between the energy and phase of these four states is precisely that of a regular SNS junction.  In particular, one obtains the ``diamond" around $\phi = \pi$, whose width is set by the Zeeman field, and gaps which depend on the rest of the parameters ($\mu, k_x, \alpha$). We also highlight in the inset, the behavior of the gaps for increasing values of $\mu$.
In this regime, as $\mu$ increases, the gaps become larger as well.
\begin{figure}
	\includegraphics[width=0.47\textwidth]{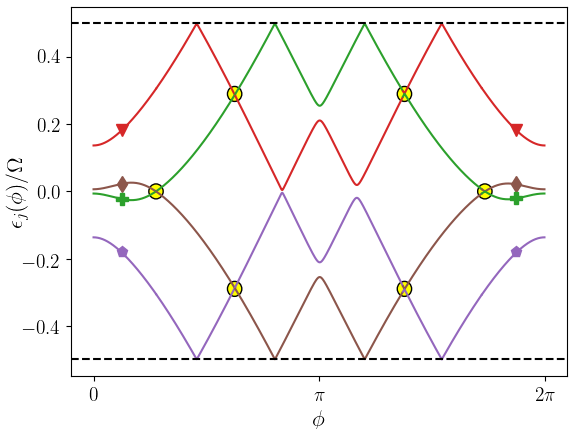}
	\caption{Floquet Andreev bound state quasienergy spectrum as a function of $\phi$. The quasienergy spectrum can be understood using the Floquet zone folding picture applied to the static spectrum of Fig.~\ref{fig:static_abs_ephi}. The crossings circled are induced by the Floquet driving and, as in the case of $\Delta \epsilon_{g,\pi}$, they have gaps which are determined by the properties of the driving. The colors and markers correspond to those from Fig.~\ref{fig:static_abs_ephi} and are helpful to illustrate the folding into the Floquet zone.}
	\label{fig:floquet_abs_qphi}
\end{figure}

The size of these gaps is relevant for identifying the periodicity as a function of $\phi$. In the special case of vanishing gaps (or when the
gaps are so small so that Landau-Zener physics is relevant), this spectrum will appear to have a period of $4\pi$ instead of $2\pi$.
However, it is generally expected that the crossings are avoided for conventional SNS junction, giving a $2\pi$ periodicity. The avoided crossings are induced by the fact that transparency of the junction is not precisely equal to $1$. In this work, the transparency is less than $1$ as we have implemented a full quantum mechanical tight-binding model which allows for, and indeed includes, scattering processes that lead to imperfect transparency in the junction.
The periodicity  for unconventional junctions (such as $p$-wave - N - $p$-wave), is on the other hand $4\pi$ \cite{Yakovenko1, Yakovenko2}. When transparency is $1$, the periodicities for conventional and unconventional junctions are both $4\pi$.

In the Floquet setting, instead of studying the relation between the energy spectrum and $\phi$, we examine how the quasienergy spectrum varies
with $\phi$. In Fig.~\ref{fig:floquet_abs_qphi}, we examine the quasienergy spectrum for the four states closest to zero quasienergy. This spectrum can be related to the static energy spectrum by considering only the four colored and marked states in Fig.~\ref{fig:static_abs_ephi}. For small amplitude driving, the quasienergy spectrum can be directly related to the energy spectrum by
folding the latter into the Floquet zone $\left[-\Omega/2, \Omega/2 \right]$. Following this, additional avoided crossings are introduced
that were not present in the static situation. Specifically, all the crossings except for the ones around the original ``diamond" are new.
The gaps at new avoided crossings induced by the Floquet folding are related to the properties of the driving, with the amplitude of the drive controlling the size of the gaps.

\section{Finite wire and experimental signatures}\label{wire_driving}
We now consider a finite system by imposing open boundary conditions instead of periodic boundary conditions.
This provides an opportunity to examine the bulk-boundary correspondence between the Floquet topological invariants and the boundary Floquet modes.
Considering a finite system also gives access to experimentally relevant quantities, such as the tunneling density of states at the boundary,
and the differential conductance.

\subsection{Mapping to one-dimensional finite wire}
We map the system to a 1D wire with $N_{1D}$ sites, $\tilde{H}_\text{BdG}\lf k\ri \to H_{1D}$, in a similar fashion to Ref.~\onlinecite{LiuShabaniMitra},
\begin{align}
H_{1D}&=\hspace{-10pt} \sum_{\substack{j = 1,\lvert i\rvert \leq N_{1D}/2,\\ 1 < j+i\leq N_{1D}}} \hspace{-12pt} t_{i} c_{j}^\dag c_{j+i} + \Delta_{i}c_{j}^\dag c_{j+i}^\dag + \text{h.c.} + \mu_{1D}n_j.
\end{align}
The parameters $\mu_{1D}, t_i, \Delta_i$ are extracted from the bulk 1D system defined by the winding angle $\alpha^k$ and $c_j^\dag (c_j)$
are creation (annihilation) operators for the 1D wire and $n_j \equiv c_j^\dag c_j$.

\begin{figure}
	\includegraphics[width=0.47\textwidth]{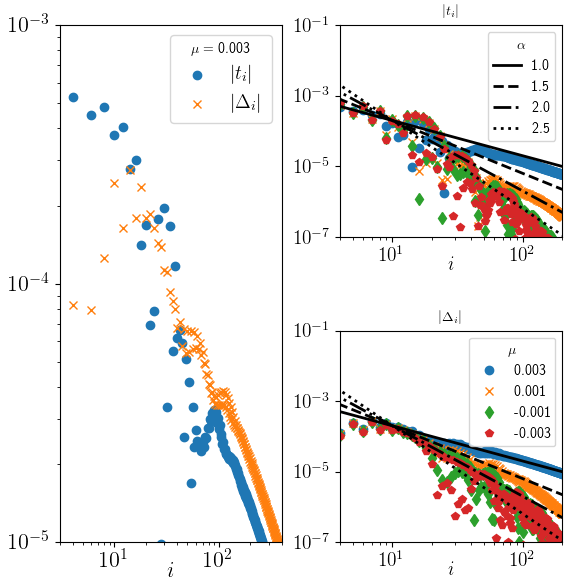}
	\caption{(left) Pairing and hopping terms for the 1D effective wire for a single value of $\mu$. These show a hybrid power-law/exponential decay as a function of separation. Hopping (top right) and pairing (bottom right) couplings for several values of $\mu$ with power-laws $\propto r^{-\alpha}$ shown, where $\alpha = 1,1.5,2,2.5$.}
	\label{fig:long_range_coupling}
\end{figure}

This mapping gives an effective wire which has substantially long-range hopping and pairing terms. In Fig.~\ref{fig:long_range_coupling}, we show these
couplings as a function of lattice site separation. The couplings decay according to a power-law $\propto r^{-\alpha}$ for an intermediate regime,
before falling off exponentially for very large separations (not shown).

We consider spatial disorder by adding $H_W$ to $H_{1D}$, where $\left[H_W\right]_{ij} = \delta_{ij}w_i$ for $w_i$ uniformly drawn from
$\left[-W,W\right]$, and for sites $i,j$ in the 1D system. Although this form of disorder is simple in the 1D effective wire language
(being uniformly distributed, random, on-site potentials), it has a complicated representation in the full 2D setting.
However, this form is sufficient to ensure the bulk states in 1D become localized, despite the long-range nature of hopping and pairing terms in
$H_{1D}$. In the numerics for the finite system, we show results for wires with length $N_{1D} = 200$. We take the disorder strength $W = 10^{-4}$,
which is smaller than the maximum $t_i, \Delta_i$, as shown in Fig.~\ref{fig:long_range_coupling}. For the differential conductance measurements shown below, we
include $200$ disorder realizations and the statistical error bars are typically smaller than the points shown.

To incorporate periodic driving in the 1D setting, we consider the static Hamiltonian for two values of $\mu$ in the normal
channel, $\mu_0$ and $\mu_0 + \delta\mu$. By mapping each of these to 1D, we then define,
\begin{align}
\delta H &\equiv H_{1D}\lf \mu_0 + \delta \mu \ri - H_{1D}\lf \mu_0\ri.
\end{align}
Above we always consider small changes in the chemical potential $\delta \mu\ll \mu_0$. The Floquet driving is introduced via
$\delta H \lf t\ri = \delta H \cos \lf \Omega t\ri$, and adopting a 1D Hamiltonian of the form
$H_{1D} \lf t; \mu_0\ri = H_{1D}\lf \mu_0\ri + \delta H\lf t\ri$. $\delta\mu$ is the same as in Eq.~\eqref{drivingterm},
but the mapping is approximate and differences between the bulk and finite Floquet systems are attributable to that. We can now find the quasienergy spectrum
and Floquet modes by diagonalizing the Sambe space representation of $H_{1D}$.

Before studying the properties of the finite wire in detail, we consider the bulk version of our 1D periodically driven system and compare
it with the discussion in Sec.~\ref{bulk_driving}. In Fig.~\ref{fig:quasienergy_spectrum}, we plot the quasienergy spectrum for both
$\tilde{H}^k_\text{BdG}$ and $H_{1D}$ with periodic boundary conditions. The spectra are in very close agreement at all points except
the most resonant $k$ points, as expected, but still differ very little. The difference is directly related to the form of driving
as implemented by $\delta H(t)$.

\subsection{Topological Floquet modes}
Once we have the Floquet modes for the 1D system, we examine the modes with quasienergy closest to $0$ and $\pm \Omega/2$.
We expect these modes to be topological in nature (i.e. robust to system size effects and disorder).
\begin{figure}
	\includegraphics[width=0.47\textwidth]{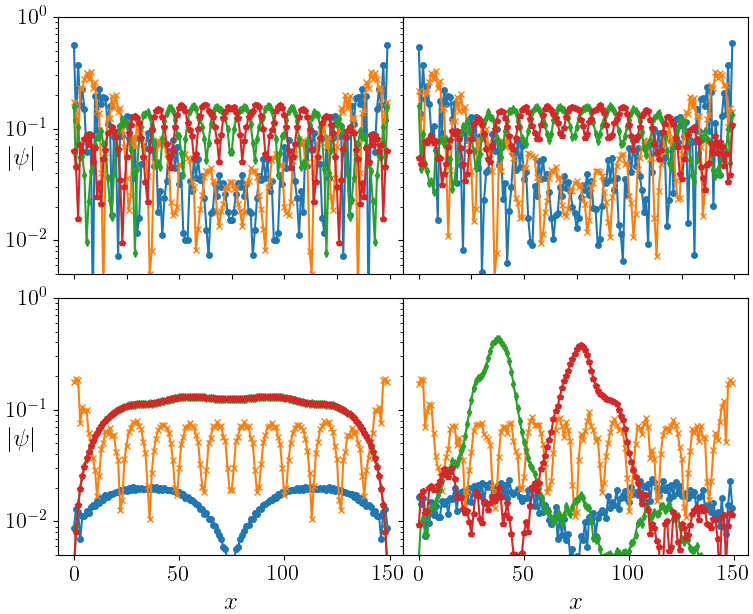}
	\caption{(top left) Amplitudes of four Floquet modes with positive quasienergy closest to $0$ for a clean system are shown. There are two edge modes
with exponential decay into the bulk and two higher energy bulk modes which are extended. (top right) Amplitudes of four Floquet modes with positive quasienergy closest to $0$ for a disordered system are shown.
Notably, all four modes are mostly unaffected by the disorder. (bottom left) Amplitudes of four Floquet modes with positive quasienergy closest
to $\Omega/2$ for a clean system are shown. There is one mode which is localized at the edge with very slow decay (orange $\times$)
and three modes extended in the system, two of which are degenerate. (bottom right) Amplitudes of four Floquet modes with positive quasienergy closest
to $\Omega/2$ for a disordered system are shown. The disorder has localized two of the extended states, but left the other two modes unaffected.
We identify the modes that are unaffected by disorder as the Majorana $\pi$ modes.}
	\label{fig:topological_modes}
\end{figure}

In the clean system, from Fig.~\ref{fig:topological_modes}, we see that there are two MZMs at the edges of the system, as expected from $\nu_0=2$ for
the chosen parameters. The situation for the MPMs is less clear, as it appears that there is only one MPM at the edge when we expect two since $\nu_{\pi}=2$.
We probe this by considering disorder, which will cause the bulk states to be Anderson localized, but leave the topological modes unscathed.

We show the modes in the presence of disorder as described above, again in Fig.~\ref{fig:topological_modes}. Here we find that the MZMs
are unsurprisingly unaffected. Strikingly, we also see that two modes with quasienergy close to $\pm \Omega/2$ are unaffected by the disorder.
These are presumably that topological MPMs. They are not well localized because their localization length is controlled by the
$\pi$ gap, which is very small. However, despite this, they are less sensitive to the inclusion of disorder, in contrast to the
bulk states which are seen to get localized.

Although the bulk states near quasienergy $\Omega/2$ become localized in the presence of disorder, it appears from
Fig.~\ref{fig:topological_modes} that the bulk states near quasienergy $0$ do not become localized. This is because the
disorder we are adding is weak as compared to the gap at the zero quasienergy, and so is a weak perturbation for all
states in the vicinity of zero quasienergy. In fact, we have checked that these states also become localized when disorder strength
and system size are increased.

\subsection{Signatures in experiments}
We now turn to experimental signatures of the zero- and $\pi$-Majorana modes. We find that there are distinctive signatures in two quantities,
the tunneling density of states and the differential conductance.
\begin{figure*}
	\includegraphics[width=0.95\textwidth]{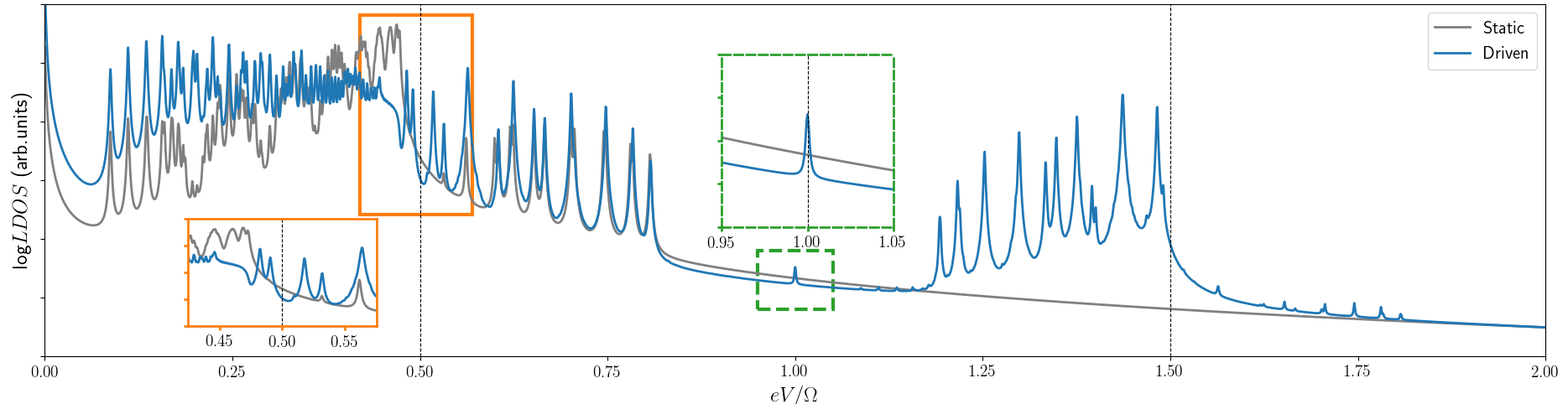}
	\caption{LDOS at edge ($x=0$) for a static (grey line) and a periodically driven (blue line) finite wire. As expected, the static case only displays features in accordance with the energy spectrum with the LDOS smoothly decreasing with increasing $eV$. In contrast, the Floquet LDOS shows the effect of driving via non-zero density of states at higher energies due to the drive. Left inset: DOS near $\Omega/2$ is dramatically modified in the driven case. Right inset: DOS near $\Omega$ is also strongly modified in the driven case. The static data are shifted to align with the driven data at large $E$. A finite Lorentzian broadening of width $10^{-6}$ has been used.}
	\label{fig:LDOS_static_comparison}
\end{figure*}

To probe these transport quantities, we use the Floquet-Landauer formalism which has been utilized previously in related topological Floquet
setups \cite{Hanggi, KunduSeradjeh, FarrellPereg-Barnea2015, FarrellPereg-Barnea2016, Yap, YapPlateau}. We attach static leads to the 1D effective wire in the wide-band
and weak-coupling limits. The assumption of equilibrium in the leads allows one to integrate them out and obtain an equation of motion for the Green's
function in the wire. Solving this equation of motion gives the Green's function, which notably inherits the time-periodicity from the periodic
Hamiltonian in this problem. For the interested reader, we refer to Appendix~\ref{FL} or the above papers for comprehensive details.

Using the Green's function, we now compute the local density of states and the differential conductance for the Floquet problem.
First, the local density of states can be computed by an expression similar to the usual one, but with additional time-averaging
\begin{align}
\overline{\rho} \lf x; E \ri & = -\frac{1}{\pi}\int_0^T \frac{dt}{T} \im G_{x,x}\lf E,t\ri,
\end{align}
where $G_{x,x}$ is the Green's function evaluated at position $x$ in the wire. The time-averaging over $t$ is necessary because we are in the
Floquet setting. The averaging smooths out the micro-motion within a drive cycle of the Floquet modes.

In the presence of a MZM, one expects the same phenomenology as in the static case: a peak at zero energy at the edge of the wire in the local density of states.
In Fig.~\ref{fig:LDOS_static_comparison}, we compare the edge local density of states for two cases, static and Floquet. In the static case, we have $\nu_0=2$,
therefore, even for the static case, there are two zero modes.

The peak at zero energy is clear in both cases, indicating the presence of MZMs. In the Floquet scenario, the presence of MPMs is expected to lead to
additional peaks at the edge of the wire at energies $\pm \Omega/2$. One also expects higher harmonic echoes of the MZM and MPM peaks at
energies $m\Omega$ and $(m\pm 1/2)\Omega$ which are generated by multiphoton processes from the driving.
These are also clearly seen in Fig.~\ref{fig:LDOS_static_comparison}.

The peaks around $\Omega/2$ are complicated by the fact that the density of states of the static system alone has weight around $E = \Omega/2$.
However, as shown in Fig.~\ref{fig:LDOS_static_comparison}, we see that in the presence of driving, the density of states around $\Omega/2$ is
significantly modified from the static case. Note that the MPMs expected for these parameters hybridize
because of the longer range interactions of the effective 1D model and so split away from each other and from precisely $\Omega/2$.
Since there are 2MPMs, their hybridization across the wire appears as 4 peaks, 2 on each side of $\Omega/2$.

\begin{figure}
	\includegraphics[width=0.47\textwidth]{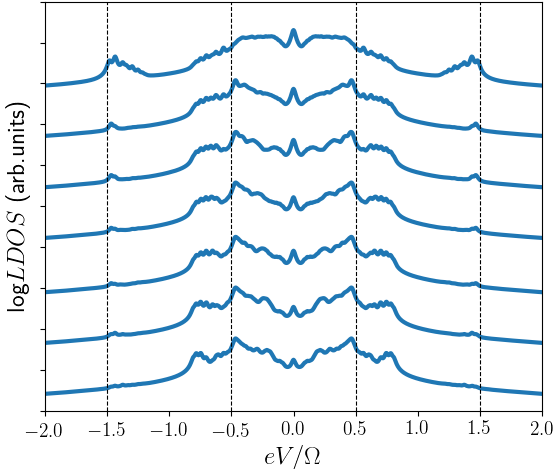}
	\caption{Local density of states for periodically driven finite wire (shown with log-scale vertical axis). Curves correspond to $\overline{\rho}\lf x;eV\ri$ for $x = 1$ to $x = 7$ from top to bottom with fixed offset for visual clarity. The peak at zero energy decays as one moves deeper into the wire. The MPM expected at energies $\pm \Omega/2$ is difficult to resolve because of contributions from the static bulk band. Higher harmonic echoes appear as peaks at energies separated from zero and $\Omega/2$ by $\Omega$. The peaks at $\pm 3\Omega/2$ are prominent, however, the peaks at $\pm\Omega$, related to the MZM, are not visible for the broadening of the Lorentzian used here for overall visual clarity. The Lorentzian broadening of width $10^{-5}$ has been introduced.}
	\label{fig:LDOS_offset}
\end{figure}

In Fig.~\ref{fig:LDOS_offset}, the MZM at the edge of the wire is identified by the peak at zero energy which decays as one moves deeper into the wire
(as seen from the diminishing peak in the offset curves corresponding to $x=1$ to $x=7$). The MPM is expected at energies $\pm \Omega/2$,
however, this is difficult to resolve because of the contributions from the static band at these energies, and its longer decay length due to
the much smaller quasienergy gap at $\Omega/2$.
This slower decay (relative to the MZM) matches the expectations from
Fig.~\ref{fig:topological_modes} that shows MPMs with much longer localization lengths than the MZMs.
The higher harmonic echoes are also detectable in the local density of states. In Fig.~\ref{fig:LDOS_offset}, these echoes appear as peaks
at energies separated from zero and $\Omega/2$ by $\Omega$. In particular, the peaks at $\pm 3\Omega/2$ are prominent, and is directly related to
the MPM. However, the peak at $\Omega$, which is related to the MZM, is not visible for the broadening of the Lorentzian used here for overall visual clarity. The peak at $\Omega$ is visible in Fig.~\ref{fig:LDOS_static_comparison}.

Differential conductance is another important experimental observable in the context of Majorana physics because of the robustly quantized
zero-bias peak that appears in the presence of a zero energy state.
In periodically driven systems, even linear response differential conductance can show additional structure beyond the zero-bias peak.
\begin{figure*}
	\includegraphics[width=0.95\textwidth, height = 0.25\textwidth]{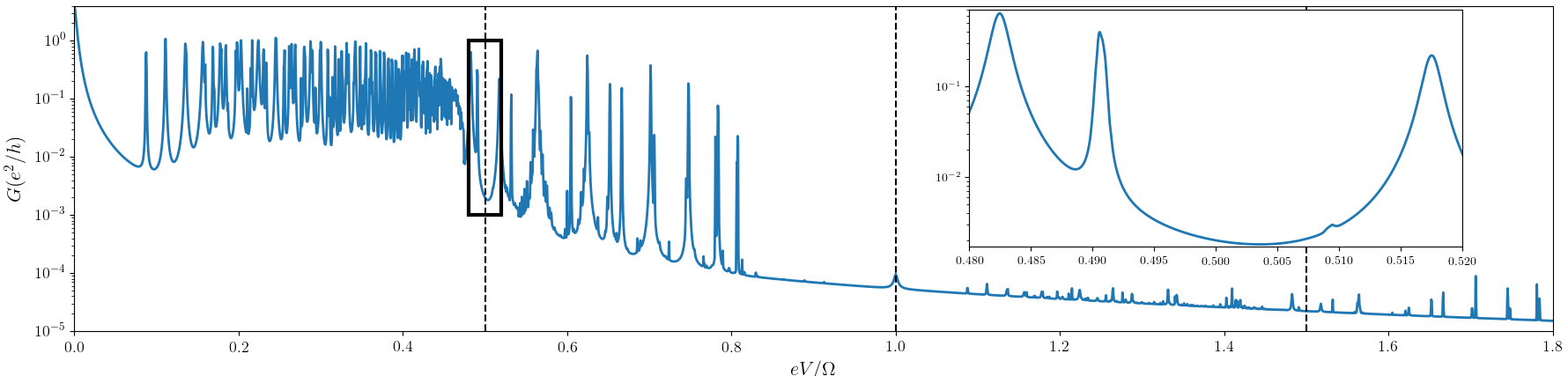}
	\caption{Physical conductance for periodically driven finite wire with disorder averaging. Inset: peaks near $\Omega/2$ corresponding to transport of MPMs are highly asymmetric in magnitude. There are 4 peaks corresponding to the 2 weakly hybridized MPMs ($\nu_{\pi}=2$) at each end of the wire. Despite the asymmetry in the magnitude of the peaks (which is in part due to the overall decay of the LDOS as a function of increasing energy), the position of the peaks remains robustly symmetric around $eV = \Omega/2$.}
	\label{fig:conductance_phys}
\end{figure*}

As with the local density of states, the computation of the differential conductance is altered slightly in the Floquet setting, by a slight
modification of the Landauer formula. The time-averaged differential conductance in the limit of zero temperature of the leads, is given by
\begin{align}
\sigma \lf E\ri &= \Gamma^2 \sum_{p={\rm int}}\lvert G^{p}_{1,2 N_{1D}}\rvert^2 +2\lvert G^{p}_{1,N_{1D}+1}\rvert^2 + \lvert G^{p}_{1,N_{1D}} \rvert^2,
\end{align}
where $\Gamma$ is the imaginary part of the self-energy that one obtains by integrating out the leads, and
$p$ denotes the Floquet harmonic. $G^p_{x,x'}$ are p-th Fourier component of
the non-equilibrium Green's function, as detailed in Appendix~\ref{FL}. The position indices refer to the different possible scattering processes
that can contribute to the conductance. The processes included are the usual Landauer term, $\lvert G^{p}_{1,N_{1D}}\rvert^2$, the
Andreev contributions at the left lead, $2\lvert G^{p}_{1,N_{1D}+1}\rvert^2$, and the right lead, $\lvert G^{p}_{1,2N_{1D}} \rvert^2$.

Unlike the static situation, in which the quantization of the zero-bias peak in the differential conductance is expected for MZMs,
the Floquet scenario requires more care. In order to recover quantization, one must consider the Floquet sum rule because carriers may absorb
or emit photons in the driven wire \cite{KunduSeradjeh, FarrellPereg-Barnea2015, FarrellPereg-Barnea2016}. The Floquet sum rule is
\begin{align}
\tilde{\sigma}\lf \epsilon \ri &= \sum_m \sigma\lf \epsilon + m\Omega\ri,
\end{align}
where $\epsilon \in \left[-\Omega/2, \Omega/2\right]$ and $\tilde{\sigma}\lf \epsilon\ri$ is the sum of the physical conductances
at $\epsilon$ shifted by $m\Omega$.
\begin{figure*}
	\includegraphics[width=0.95\textwidth, height = 0.25\textwidth]{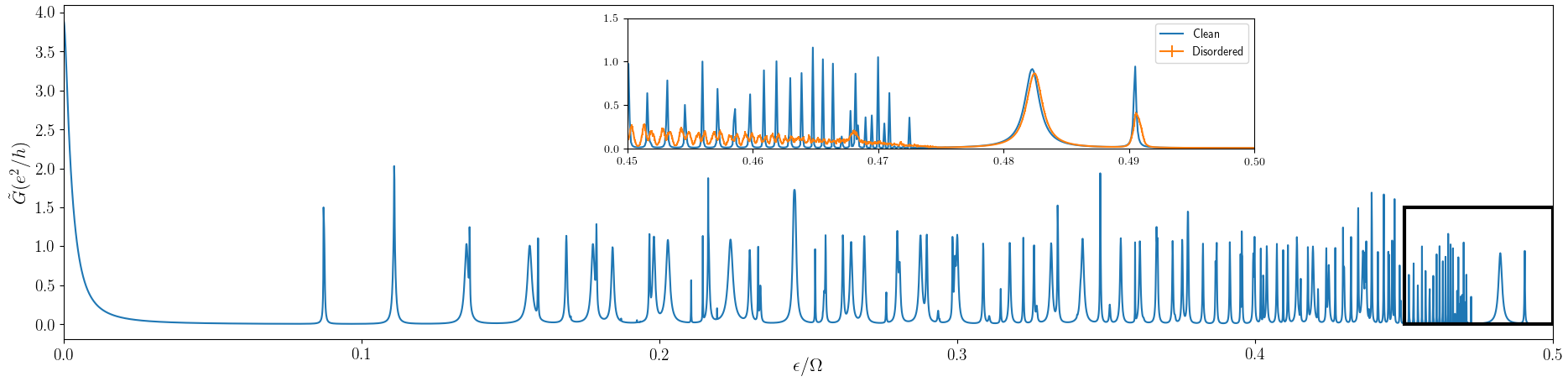}
	\caption{Sum-rule conductance for the periodically driven clean finite wire. Inset: comparing clean and disorder averaged wires
shows a significant reduction for the peaks for the latter away from the $\Omega/2$ Floquet zone boundary. Thus disorder helps to better identify
the MPMs.}
	\label{fig:conductance_sum}
\end{figure*}

Quantization aside, the zero-bias and Floquet MPM peaks at biases $\pm \Omega/2$, will still appear in the differential conductance.
However, the formerly quantized height of these peaks may be distributed to biases separated by integer multiples of the drive frequency,
$m\Omega$.

In Fig.~\ref{fig:conductance_phys}, the differential conductance as a function of voltage bias is shown for a wide range on a log-scale.
The zero-bias peak is clearly visible and in fact reaches nearly $4e^2/h$, as expected for a system with $2$ MZMs.
There is an additional peak visible at the first integer multiple of the drive frequency, as expected from the above discussion
about the sum-rule. On each side of the bias corresponding to $\pm \Omega/2$, there are two peaks, including one which is particularly
diminished, but still visible, at $eV \approx 0.51\Omega$. These peaks correspond to the presence of MPMs, of which we expect $2$
to be present for the chosen parameters.
Their quantization is much weaker than that of the MZM peaks.  The long decay lengths cause the MPMs to be split slightly, and to also be diminished
in height.

Lastly, in Fig.~\ref{fig:conductance_sum}, we have computed the sum-rule conductance. In terms of this quantity, the MZMs remain clearly visible
at the zero-bias peak.
However, the MPMs are strikingly more evident than in the regular conductance.
In particular, when one considers disorder averaging (see inset of Fig.~\ref{fig:conductance_sum}), the peaks corresponding to MPMs
acquire a stronger peak height. It is precisely this effect that Ref.~\onlinecite{KunduSeradjeh} refer to when they suggest that disorder
may be used to aid in the identification of MPMs.

The presence of both MZMs and MPMs in a system gives rise to ``Floquet time-crystal" dynamical effects, such as period doubling, or equivalently,
subharmonic responses in observables. In the experimentally relevant observables we have calculated and described above,
this phenomenon is present and visible as peaks that are spaced apart by $\Omega/2$ in energy, rather than $\Omega$.

\section{Summary and outlook} \label{summary}
In the present work, we have shown how one can combine several developments in theory and experiment, to realize Floquet Majorana fermions
in a mesoscopic system. In particular we have outlined how one can model the Floquet Andreev bound state modes of a periodically
driven planar Josephson junction,
as a 1D Kitaev chain with long range couplings and periodic driving. Then we have discussed the properties of the Floquet
topological modes, highlighting their distinctive signatures in local density of states, differential conductance measurements,
and discussed their robustness to disorder.

In terms of future directions, it would be interesting to investigate the behavior of real-space topological invariants for these systems along the lines
of Ref.~\onlinecite{LiuShabaniMitra}. Constructing real-space topological invariants \cite{HastingsLoringEPL, Loring} for Floquet systems
is also an interesting mathematical question.

Another important remaining question is that of realistic disorder. We have considered disorder which is straightforward in the 1D wire,
but which is unlikely to have a simple representation in the physical system.
Instead, one could consider a random matrix approach to the disorder in the presence of Floquet driving \cite{Movassagh}.

Recently it has been argued that not only Majorana fermions in nanowires, but also coalescing Andreev modes in a topologically trivial regime, can show similar experimental signatures \cite{CXLiuABS}. The planar Josephson junctions we study are different from the nanowire geometry and have shown great promise for realization of Majorana fermions over a wide range of parameter space \cite{YacobyPJJexpt, MarcusPJJexpt}. Our work probes a new dimension in topological superconductivity through the addition of periodic driving. This non-equilibrium probe may help shed light on the types of questions raised in nanowires.

We note that experimental progress with these semiconductor-superconductor heterostructures is proceeding rapidly. In addition to
direct transport and tunneling measurements that we have presented here, a study of the effect of Floquet periodic driving on
Shapiro steps could yield similarly distinctive signatures of the topological Floquet physics.
Including strong interactions \cite{InteractingInAs}, and thus identifying an experimental regime for
Floquet parafermions \cite{FloquetPara}, is also an important direction for research.
Lastly, we note that many recent ideas have suggested ways of using Floquet physics to enhance or realize necessary constituents for quantum computation, including qubits, gates, and braiding \cite{FloquetQubit, FloquetGates,FloquetQC, FloquetBraiding}, and the system studied here may be ideally positioned to realize these features.

\subsection{Acknowledgments}
We thank Yonah Lemonik and Daniel Yates for helpful discussions. DTL and AM were supported by the US Department of Energy, Office of Science,
Basic Energy Sciences, under Award No.~DE-SC0010821. JS was supported by the US Army Office of Research and US Air Force Office of Scientific Research Young Investigator Award.

\begin{appendix}
\section{MPM quasienergy gap}\label{RWA}
In this appendix, we outline a simple approach to estimate the gap corresponding to the MPMs.
We consider a two-level system near resonance. Specifically, we consider the Hamiltonian
\begin{align}
H\lf t\ri &= \lf\begin{matrix}
\epsilon_0 && 0 \\
0 && -\epsilon_0
\end{matrix}\ri + \lf \alpha_x \sigma_x + \alpha_y \sigma_y\ri \cos \Omega t.
\end{align}

This gives the following Schr\"odinger equation,
\begin{align}
i\partial_t \lf\begin{matrix} \psi_1 \\ \psi_2
\end{matrix}\ri &=
\lf \begin{matrix}
\epsilon_0 && \alpha_+  \cos \Omega t\\
\alpha_-  \cos \Omega t && -\epsilon_0
\end{matrix}\ri \lf \begin{matrix} \psi_1 \\ \psi_2
\end{matrix}\ri,
\end{align}
where $\alpha_\pm = \alpha_x\pm i \alpha_y.$ Now we go to a rotating frame
\begin{align}
\lf \begin{matrix} \psi_1 \\ \psi_2
\end{matrix}\ri &= \lf\begin{matrix} e^{-i\lf\epsilon_0 + \Delta/2\ri t} \tilde{\psi}_1 \\   e^{i\lf\epsilon_0 + \Delta/2\ri t}\tilde{\psi}_2
\end{matrix}\ri,
\end{align}
where $\Delta \equiv \Omega - 2\epsilon_0$. Upon neglecting terms which rotate quickly (those proportional to $e^{\pm2i\Omega t}$),
we have the following time-independent problem
\begin{align}\label{star}
i\partial_t \lf\begin{matrix} \tilde{\psi}_1 \\ \tilde{\psi}_2
\end{matrix}\ri &=
\lf \begin{matrix}
-\Delta/2 && \alpha_+/2 \\
\alpha_-/2 && \Delta/2
\end{matrix}\ri \lf \begin{matrix} \tilde{\psi}_1 \\ \tilde{\psi}_2
\end{matrix}\ri.
\end{align}

This is solved by
\begin{align}
\lf \begin{matrix} \tilde{\psi}_1 \\ \tilde\psi_2
\end{matrix}\ri &=  e^{-i\epsilon_k t}\lf\begin{matrix} \phi_1 \\ \phi_2
\end{matrix}\ri,\\
\epsilon_k &= \pm \frac{1}{2} \sqrt{\Delta^2+\alpha_x^2+\alpha_y^2}.
\end{align}

To fully connect this approach to the problem we consider in the main text, we must also include a term $\sigma_z \alpha_z \cos\Omega t$. This term is unaffected by the rotating frame transformations. Consequently, Eqn.~\eqref{star} becomes
\begin{align}\label{twolevelzcomp}
i\partial_t \lf\begin{matrix} \tilde{\psi}_1 \\ \tilde{\psi}_2
\end{matrix}\ri &=
\lf \begin{matrix}
-\Delta/2 + \alpha_z \cos\Omega t && \alpha_+/2 \\
\alpha_-/2 && \Delta/2  - \alpha_z \cos\Omega t
\end{matrix}\ri \lf \begin{matrix} \tilde{\psi}_1 \\ \tilde{\psi}_2
\end{matrix}\ri.
\end{align}

We now perform another unitary transformation, $U$
\begin{align}
U\lf t\ri &= \lf\begin{matrix} e^{-i\Delta t/2} e^{i\alpha_z\int^t \cos\Omega t' dt'} && 0\\
0 && e^{i\Delta t/2}  e^{-i\alpha_z\int^t \cos\Omega t' dt'}\end{matrix}\ri.
\end{align}

Using the standard Bessel function identity, we expand
\begin{align}
e^{-i\Delta t}e^{\frac{2i\alpha_z}{\Omega}\sin\Omega t} &= e^{-i\Delta t} \sum_m e^{i m\Omega t}J_m\lf \frac{2\alpha_z}{\Omega}\ri.
\end{align}

Since the amplitude $\Delta$ is small, we keep only the term $m=0$. Hence, Eqn.~\eqref{twolevelzcomp} becomes
\begin{align}\label{twolevelzcomp2}
i\partial_t \lf\begin{matrix} \tilde{\psi}_1 \\ \tilde{\psi}_2
\end{matrix}\ri &=
\lf\begin{matrix} 0 && \frac{\alpha_+}{2} e^{-i\Delta t} J_0 \lf \frac{2\alpha_z}{\Omega}\ri  \\
  \frac{\alpha_-}{2} e^{i\Delta t} J_0 \lf \frac{2\alpha_z}{\Omega}\ri && 0\end{matrix}\ri
  \lf \begin{matrix} \tilde{\psi}_1 \\ \tilde{\psi}_2
\end{matrix}\ri.
\end{align}
Again we go to a rotating frame
\begin{align}
\lf \begin{matrix} \tilde\psi_1 \\ \tilde\psi_2
\end{matrix}\ri &= \lf\begin{matrix} e^{-i\frac{\Delta}{2} t} \phi_1 \\   e^{i\frac{\Delta}{2} t}\phi_2
\end{matrix}\ri,
\end{align}
and Eqn.~\eqref{twolevelzcomp2} becomes
\begin{align}
i\partial_t \lf\begin{matrix} {\phi}_1 \\ {\phi}_2
\end{matrix}\ri &=
\lf\begin{matrix} -\frac{\Delta}{2} && \frac{\alpha_+}{2}J_0 \lf \frac{2\alpha_z}{\Omega}\ri  \\
  \frac{\alpha_-}{2}J_0 \lf \frac{2\alpha_z}{\Omega}\ri && \frac{\Delta}{2}\end{matrix}\ri
  \lf \begin{matrix} {\phi}_1 \\ {\phi}_2
\end{matrix}\ri,
\end{align}
which has the following eigenvalues
\begin{align}
\tilde{\epsilon}_k &= \frac{1}{2}\sqrt{\Delta^2 + J_0^2\lf \frac{2\alpha_z}{\Omega}\ri \lf \alpha_x^2 + \alpha_y^2\ri}.
\end{align}

To make a concrete connection with the full problem of the main text, we use $\epsilon_0 \equiv E_0$.
Then, to define $\alpha_x,\alpha_y, \alpha_z$, we start from the $2\times2$ matrix representation of $\tilde{H}_d$ (obtained after projecting to the sub-gap bands). This $2\times2$ matrix defines components for $\sigma_x,\sigma_y,\sigma_z$, from which we extract the coefficients $\alpha_x,\alpha_y,\alpha_z$.

\section{Floquet-Landauer formalism}\label{FL}
Here we give very basic background on the method used to compute the non-equilibrium Green's function and transport-related quantities, following closely the appendix of Ref.~\onlinecite{FarrellPereg-Barnea2016}. For a more expansive and detailed description, see further Refs.~\onlinecite{Hanggi, KunduSeradjeh, FarrellPereg-Barnea2015}.

We take a one-dimensional tight-binding Hamiltonian, such as $H_{1D}$ from the main text. We couple this to leads, $L$ and $R$:
\begin{align}
H\lf t \ri &= H_{1D}\lf t\ri + H_\text{leads} + H_{LW},\\
H_\text{leads} &= \epsilon_L c_L^\dag c_L + \epsilon_R c_R^\dag c_R,\\
H_{LW} &= \lf c_L\, c_L^\dag \ri V_L \lf \begin{matrix}c_1^\dag \\ c_1 \end{matrix}\ri+\lf c_R\, c_R^\dag \ri {\cal V}_R \lf \begin{matrix}c_{N_{1D}}^\dag \\ c_{N_{1D}} \end{matrix}\ri.
\end{align}

Using this setup, we solve for the Green's function, $G\lf E, t \ri$, which satisfies the following time-dependent Schr\"odinger equation
\begin{align}
\lf i\frac{d}{dt} + E - H_{1D}\lf t\ri \ri G\lf E, t\ri  \nonumber \\
\quad +i\int^\infty_{-\infty}e^{iEt'} \Gamma\lf t'\ri G\lf E, t - t' \ri dt' = \mathbb{1},
\end{align}
where the above is a $2N_{1D}\times 2N_{1D}$ matrix equation (the factor of $2$ arises because we are working with a Bogoliubov-de Gennes formalism). $\Gamma$ is the imaginary part of the self-energy obtained by integrating out the leads. Explicitly, it can be written as
\begin{align}
\Gamma\lf t\ri &\equiv \Gamma_L + \Gamma_R = {\cal V}_L^\dag g_L {\cal V}_L + {\cal V}_R^\dag g_R {\cal V}_R,
\end{align}
where $g_{L,R}\lf t\ri$ is the Green's function for the leads, which will be taken in the wide-band limit.
In particular, after Fourier transforming, the imaginary part of the self-energy is $\im \Sigma\equiv \Gamma = \pi \rho_\text{lead}v^2$,
where $\rho_\text{lead}$ is the density of states in the leads (a constant in the wide-band limit)
and $v$ is the coupling between the leads and the wire. The leads are assumed to have Fermi-Dirac distributions,
$f_{L,R}\lf E\ri = 1/\lf1+ e^{\beta\lf E-eV_{L,R}\ri} \ri,$ for inverse temperature $\beta$ and voltage bias $V_{L,R}$.
For simplicity, we take the voltage bias $V_L = V$ and $V_R = 0$.

The Green's function can now also be expanded in a Fourier series. This expansion is similar to the Sambe space representation of the
Hamiltonian and Floquet modes. The object obtained, $G^p \lf E\ri$ is the key computational building block
for the local density of states and differential conductance for Floquet systems, and has the form,
\begin{align}\label{FMexpn}
    G^p\lf E\ri &= \frac{1}{T}\int_0^T dt e^{ip\Omega t}G\lf E, t\ri,\\
    G^{p} \lf E \ri &= \sum_{\alpha,m} \frac{\lvert\phi_\alpha^{p+m}\rangle\langle\phi_\alpha^{m}\rvert}{E- \lf \epsilon_\alpha - m\Omega -i\gamma_\alpha\ri},\\
    \gamma_\alpha &= \int_0^T\frac{dt}{T}\langle\phi_\alpha\lf t\ri\lvert\Gamma \rvert\phi_\alpha\lf t\ri \rangle,
\end{align}
where we take the weak-coupling limit so that the leads may be treated perturbatively and the Floquet quasienergies are not changed to first order, but the Floquet modes are modified. Now we can, for example, compute the differential conductance using the Floquet-Landauer formula for the time-averaged current at zero temperature, $\sigma \lf E\ri = e\frac{\partial \overline{I}}{\partial V}\rvert_{V=0, T\to0}$:
\begin{widetext}
\begin{align}
\sigma \lf E\ri &= \frac{\partial}{\partial V} \sum_p\int dE \Tr \left[ \Gamma_L G^p \Gamma_R G^{p \dag} \right]\lf f_R\lf E\ri - f_L\lf E\ri \ri + \Tr \left[ \Gamma_L G^p \Gamma_R G^{p \dag}\right]\lf 1 - f_R\lf -E \ri - f_L\lf E\ri\ri\\
&= \lf\pi \rho_\text{lead} v^2\ri^2 \sum_{p}\lvert G^{p}_{1,2 N_{1D}}\rvert^2 +2\lvert G^{p}_{1,N_{1D}+1}\rvert^2 + \lvert G^{p}_{1,N_{1D}} \rvert^2\\
   &=\lf\pi \rho_\text{lead} v^2\ri^2 \sum_{p,m,\alpha}\frac{\lvert \phi^{p+m}_{\alpha,1}\rvert^2\lvert \phi^{p}_{\alpha,2 N_{1D}}\rvert^2 +2\lvert \phi^{p+m}_{\alpha,1}\rvert^2\lvert \phi^{p}_{\alpha,N_{1D}+1}\rvert^2 + \lvert \phi^{p+m}_{\alpha,1} \rvert^2\lvert \phi^{p}_{\alpha,N_{1D}} \rvert^2}{\lf E - \epsilon_\alpha + p\Omega \ri^2 + \gamma_\alpha^2}.
\end{align}
\end{widetext}
To go from the second to third lines above, we have used the expansion of Eqn.~\eqref{FMexpn}.
\end{appendix}

\bibliography{bib.bib}
\end{document}